\documentclass[12pt]{article}

\usepackage{fancyhdr}

\usepackage{xcolor}         % colors
\usepackage{latexsym}
\usepackage{amsmath}
\usepackage{amssymb}
\usepackage{times}
\usepackage{graphicx}
\usepackage{epsfig}
\usepackage{array}
\usepackage{flafter}
\usepackage[page]{appendix}
\usepackage{listings}
\usepackage{algorithm,algpseudocode}
\usepackage{ulem}
\usepackage{verbatim}
\usepackage{hyperref}
\usepackage{qtree}
\usepackage{mathrsfs,amsmath} 
\usepackage{subcaption}
\usepackage{graphicx}

\usepackage[a4paper, total={6in, 8.5in}]{geometry}

\graphicspath{ {Figures/} }

\usepackage{amsthm}
\newtheorem*{lemma}{Lemma}
\newtheorem{ExampleDef}{Example}%[section]

\begin{document}
 
\thispagestyle{fancy}
\fancyhf{} % Clear all headers and footers
\renewcommand{\headrulewidth}{0pt} % Remove the top line
\fancyfoot[L]{\it \\ \ \ \\ This article is distributed under the terms and conditions of the Creative Commons Attribution (CC BY 4.0) license.} % Left footer
%\fancyfoot[C]{\thepage} 

\begin{center}
{\Large \bf Reply to second comment arXiv:2606.05197 on \\ "Computing quantum waves exactly from classical action" \par} \vspace{1.5em}
{\large Winfried Lohmiller and Jean-Jacques Slotine \par}
{Nonlinear Systems Laboratory \\
Massachusetts Institute of Technology \\
Cambridge, Massachusetts, 02139, USA\\
{\sl \{wslohmil, jjs\}@mit.edu} \par}
\end{center}

%The new post~\cite{vattaycomment2} also states incorrectly

\begin{abstract}

Comment on arXiv:2605.02621 \cite{Vattay2026comment} and arXiv:2606.05197 \cite{vattaycomment2}:

In contrast to his earlier comment \cite{Vattay2026comment} on rspa.2025.0413 \cite{LohmillerSlotine2026QuantumWaves} and answered in arXiv:2605.20443 \cite{LohmillerSlotine2026BohmPotential}, the same author of the second post \cite{vattaycomment2} does not seem to dispute any longer that the use in \cite{LohmillerSlotine2026QuantumWaves} of tools from Feynman's \cite{Feynman}, namely
\begin{itemize}
    \item the kernel approach, which computes a wave starting at a given initial position or momentum, and is a key element in \cite{LohmillerSlotine2026QuantumWaves} to ensure that the Bohm quantum potential vanishes. This is directly verified in all examples of \cite{LohmillerSlotine2026QuantumWaves} as well as in the Pauli, Dirac and Maxwell derivations from classical action. This approach is very different from the standard Bohm-Madelung potential on the overall wave, as illustrated in detail in \cite{LohmillerSlotine2026BohmPotential} for the two-slit experiment.
    \item  for the harmonic oscillator, the kernel computation directly from a Taylor series expansion, which does not involve any circularity. 
\end{itemize}
His second post~\cite{vattaycomment2} now
\begin{itemize}
    \item criticizes the use in \cite{LohmillerSlotine2026QuantumWaves}, \cite{LohmillerSlotine2026BohmPotential} of Duru and Kleinert's time-independent eigenwave equivalence of the original and the time-scaled Schrödinger equations~\cite{Duru}, apparently confusing this standard result with a  multivariable chain rule in the time-dependent kernel computation from action in \cite{LohmillerSlotine2026QuantumWaves}. 
    %Hence we summarize an operator-level proof \cite{{lohmiller2026exactv1} \cite{Lovelock} which shows this equivalence using a harmonic coordinate transformation. 
    \item claims that "The mathematical framework presented by the authors remains identical to the well-established semiclassical Van Vleck propagator [sic]" which requires that the "Hamiltonian is at most quadratic in position and momentum [sic]". This confuses a quadratic action in the vector case with a harmonic propagated density based on the Laplacian of the action (with harmonicity always fulfilled e.g., for holomorphic complex functions \cite{lohmiller2026exactv1}).  In addition, it ignores the extension of a real position to a complex or quaternion position or spin, the key result on multi-valued action branches (which enables a classical derivation of quantum randomness) and the quantization of classical action in \cite{LohmillerSlotine2026QuantumWaves}.
\end{itemize}

%In addition, while the first post~\cite{Vattay2026comment} contradicts a standard result of Van Vleck and Feynman~\cite{Vattay2026comment},  "For bound states with spatially varying densities -- such as the harmonic oscillator (example 3.9) [...] \ the quantum potential is manifestly non-zero [sic]",  the same author's second post~\cite{vattaycomment2} now emphasizes that result.

\noindent Keywords $-$  Response to {\it arXiv:2605.02621} and {\it  arXiv:2606.05197} in relation to {\it rspa.2025.0413}
"On computing quantum waves exactly from classical action", Feynman path integrals, Duru-Kleinert time rescaling, Bohm-Madelung potential, claimed multivariable chain rule violation in {\it  arXiv:2606.05197}

\end{abstract}

\section{Introduction}

The point that the harmonicity assumption on the {\it propagated} classical square root density in~\cite{LohmillerSlotine2026QuantumWaves} should be highlighted is well-taken and has been made before, as detailed in~\cite{LohmillerSlotine2026BohmPotential} in response to the post~\cite{Vattay2026comment}.
%The point from \cite{Vattay2026comment} that the propagated classical square root density is harmonic is well-taken and was highlighted in Lemma 3.1 in~\cite{LohmillerSlotine2026QuantumWaves} or \cite{LohmillerSlotine2026BohmPotential}.
Satisfying this assumption for general nonlinear potentials may require a time rescaling, as in the case of the hydrogen atom example in~\cite{LohmillerSlotine2026QuantumWaves}. The second post \cite{vattaycomment2} from the same author, now commenting on~\cite{LohmillerSlotine2026BohmPotential},  does not seem to dispute any longer that the use in \cite{LohmillerSlotine2026QuantumWaves} of tools from Feynman's \cite{Feynman}, namely
\begin{itemize}
    \item the kernel approach, which computes a wave starting at a given initial position or momentum, and is a key novel element in \cite{LohmillerSlotine2026QuantumWaves} to ensure that the Bohm quantum potential vanishes. This is directly verified in all examples of \cite{LohmillerSlotine2026QuantumWaves} as well as in the Pauli, Dirac and Maxwell derivations from classical action. This approach is very different from the standard Bohm Madelung potential, based the Madelung density \cite{Madelung1927} 
\begin{equation}
    \rho = |\psi|^2 \ \ \ \ \ \ \ \ \ \ \ \ \phi = \hbar\,\arg(\psi) \nonumber
\end{equation}
which, in contrast to  \cite{LohmillerSlotine2026QuantumWaves}, generates for most waves a Bohm quantum potential \cite{Bohm1952I, Bohm1952II} since the Madelung density is typically position-dependent. %This point is illustrated in detail in \cite{LohmillerSlotine2026BohmPotential} for the two-slit experiment.

    \item  for the harmonic oscillator, the kernel computation direcly from a Taylor series expansion, which does not involve any circularity. 
\end{itemize}
%\wl{which means that both positions have converged to some extent. This convergence was achieved thanks to the second part of this post which was the v3 of \cite{LohmillerSlotine2026BohmPotential}.} 
Indeed Bohm-like terms along individual stationary action paths are very different from the standard Bohm-Madelung potential computed from the overall wave. This is illust,rated for instance in the detailed double slit example in~\cite{LohmillerSlotine2026BohmPotential}, whose Bohm-like terms are exactly zero  while the usual Bohm-Madelung potential can be very large.  

We emphasize again that our analysis is based strictly on the construction of the single-point quantum propagator $K(q',t';q_0,t_0)$. The vanishing of the corresponding propagated quantum potential (which we call in~\cite{LohmillerSlotine2026BohmPotential} the Bohm-like propagated potential) is a mathematical property of the point-source kernel's density distribution \(\rho_K\), ensuring that individual classical extremal action paths directly generate the correct path-integral weight. Once kernels are integrated over initial distributions \(\Psi_0(q_0)\) to form a general wavefunction \(\Psi(q,t)\), the resulting state naturally exhibits standard quantum interference and non-zero overall Bohm quantum potential. This is one of the key novelties of \cite{LohmillerSlotine2026QuantumWaves}. Our framework neither modifies standard Feynman-kernel integration, nor claims to eliminate nonlocal quantum-potential dynamics for the fully superposed wave function, but simply computes that wave function in a different way, where Bohm-like terms from quantum propagators effectively vanish.

The same author's second post~\cite{vattaycomment2} now concentrates on the claim in~\cite{LohmillerSlotine2026BohmPotential}  that
the harmonicity assumption on the propagated classical square root density,
if not directly verified, can be achieved through time-rescaling. Specifically, \cite{vattaycomment2} states
\begin{itemize}
    \item a chain-rule objection on the equivalence of the original and  the time-scaled Schrödinger equations, "A rigorous mathematical examination of this transformation reveals a violation of the multivariable chain rule [sic]” . This confuses the time-dependent kernel computation from action in Lohmiller and Slotine \cite{LohmillerSlotine2026QuantumWaves} with the Duru and Kleinert's time-independent eigenwave equivalence of the original Schrödinger equation with the time-scaled Schrödinger equation (16) in~\cite{Duru}. This confusion of decoupled results will be detailed in Sections \ref{scaling} and \ref{eigenwave} of this post. The Duru and Kleinert approach itself was already strictly used in the original article~\cite{LohmillerSlotine2026QuantumWaves}
    
    \item "The mathematical framework presented by the authors remains identical to the well-established semiclassical Van Vleck propagator [sic]". Section \ref{Vleck} here shows that this confuses the Hessian of the action with the Laplacian of the propagated density or action, ignores the extension of a real position to a complex or quaternion spin, and finally ignores the multi-valued action branches and classical quantization in \cite{LohmillerSlotine2026BohmPotential}. 
\end{itemize}

Finally, note the vast majority of the article~\cite{LohmillerSlotine2026QuantumWaves} is decoupled from this discussion, since in all cases the provided classical density shows that the Bohm-like quantum potential vanishes along paths. Indeed, using as \cite{LohmillerSlotine2026QuantumWaves} a two-step construction based on first computing Feynman propagators (kernels) from given initial conditions, and then integrating over initial conditions, the Bohm-like quantum potentials are exactly zero in
\begin{itemize}
    \item Theorem 2.4, on branch points for the classical action
    \item Lemma 3.3, which finds at the classical branch points classical randomness which explains the randomness of quantum waves
    \item Lemma 3.4, which performs a classical quantization
    \item All examples, such as the as double slit, Aharonov-Bohm, particle in the box, tunneling, harmonic oscillator, hydrogen, EPR, and QED
    \item Section 4,  which extends the approach to Klein-Gordon, Pauli, Dirac and Maxwell equations.
\end{itemize}
posts \cite{Vattay2026comment} and \cite{vattaycomment2} ignore these main contributions of the article \cite{LohmillerSlotine2026QuantumWaves}.
%, and incorrectly equate it to a semi-classical result neglecting the standard Bohm quantum potential of the overall wave.

We certainly agree with the point in~\cite{vattaycomment2}, "The Schr\"odinger  equation is formulated in physical space  $x$. To determine if a proposed wave function constitutes an exact solution, all differential operators must be evaluated with respect to these physical coordinates". 

However one of the main messages of the abstract  \cite{vattaycomment2} "A rigorous mathematical examination of this transformation reveals a violation of the multivariable chain rule [sic]” 
\begin{equation}
    \nabla_{q} \rho \bigl(t_{j}'(q,t)\bigr) = \frac{d \rho}{dt_{j}'}\,\nabla_{q} t_{j'} \label{eq:vattayerror}
\end{equation}
claims that the chain rule on $t_{j}'(x,t)$ above was missed in \cite{LohmillerSlotine2026QuantumWaves} or \cite{LohmillerSlotine2026BohmPotential}. 

While we certainly believe in the chain rule, this is again a mis-representation of ~\cite{LohmillerSlotine2026QuantumWaves} and~\cite{LohmillerSlotine2026BohmPotential}, which do not use this approach at all. The rescaled time $t'$  is only used for mapping the action of the Hamilton-Jacobi in $t'$ into the kernel of the Schr\"odinger in $t'$ in (4.3), (4.4) and (4.5) in the appendix. The original time $t$ is not used in this mapping.
This is summarized in section \ref{scaling}. 

Section \ref{eigenwave} then shows that the eigenwave of the original and time-scaled eigenwave  are identical, i.e. it shows that actually the wave of the original Schrödinger equation was computed.

\section{Time scaled kernel in $t'$ time and stationary eigenwave } \label{scaling}

The exact computation of the quantum kernel (4.5), fulfilling the Schr\"odinger equation (4.4) from the action (4.3) in scaled time $t'$ is e.g. shown in the appendix \cite{LohmillerSlotine2026BohmPotential}.  We believe that \cite{vattaycomment2} accepts this result:

More specifcically the time scaled Schr\"odinger equation (\ref{eq:Schroedtime}) can be solved on each stationary branch $j \in \mathbb{B}$  by the propagator or Feynman kernel \cite{Feynman} 
\begin{eqnarray}
    K({\bf q}', t', {\bf q}_o \veebar {\bf p}_o) \ &=& \sum_{j \in \mathbb{B}} \sqrt{  \rho_j} \ e^{\frac{i}{\hbar} \phi_j} \ \ \ \ \ \label{eq:propagator} 
    %&=& \jjs{  \sum_k \Psi'_k({\bf q}') \Psi_k^{'*}({\bf q}_o'({\bf q}_o)) e^{-\frac{i}{\hbar} E_k' t' } } \nonumber
    %&=& \sum_{j \in \mathbb{B}} e^{- \frac{1}{2} \int_{0}^{t'} (T \Delta_{{\bf M}} \phi_j) (\theta') \ d \theta' }  e^{\frac{i }{\hbar} \phi_j} \nonumber 
\end{eqnarray}
 using the action field $\phi_j({\bf q}, t, {\bf q}_o \veebar {\bf p}_o)$ and the classical propagated harmonic square density $\sqrt{\rho_j}(t')$ using the proofs of \cite{LohmillerSlotine2026QuantumWaves} or \cite{LohmillerSlotine2026BohmPotential}. No Bohm-like quantum potential is needed due to the harmonic propagated square root density. 

Weighting and integrating the propagator (\ref{eq:propagator}) with the initial wave $\psi_{o} = \sqrt{  \rho_j }  e^{\frac{i }{\hbar} \phi_j}({\bf q}, t=0)$ leads to the general eigenwave \cite{Feynman} of the time-scaled stationary Schr\"odinger equation (\ref{eq:Schroedtime}) 
\begin{equation}
\Psi ({\bf q}) = \int_{{\bf q}_o \ \veebar \ {\bf p}_o \in \mathbb{G}^N} \Psi_{o} \ K \ d{\bf q} \ \veebar \ d{\bf p} \label{eq:Wave}
\end{equation}
if the eigen-energy offset $E_k'$ is subtracted from  $V / T$. 

\section{Original wave and stationary eigenwave} \label{eigenwave}

Instead the Duru and Kleinert time-scaling~ principle (16) in \cite{Duru} is used to map the {\it time-independent  eigenwave} $\Psi_k({\bf q}) = \Psi_k'({\bf q}'({\bf q}))$ of the original stationary Schrödinger equation to the stationary time-scaled Schrödinger equation 
%$\psi = \sum_{k} \Psi_{k}(\mathbf{q})\, e^{\frac{i}{\hbar} E_{k} t}, \psi' = \sum_{k} \Psi'_{k}(\mathbf{q}')\, e^{\frac{i}{\hbar} E_{k}' t'}$ and 
\begin{eqnarray}
\frac{1}{T }\left[ \frac{1}{2}\left( \frac{\hbar}{i} \nabla_{\bf M} - {\bf Q} {\bf A} \right) \cdot {\bf M}^{-1} \left( \frac{\hbar}{i} \nabla -  {\bf Q}  {\bf A} \right) + V - E_k \right] \Psi_k &&  \ \ \ \ \ \label{eq:SchroedCons} \\
= \ \left[ \frac{1}{2} \left( \frac{\hbar}{i} \nabla' - {\bf Q} {\bf A}' \right) \cdot   \left( \frac{\hbar}{i} \nabla' -  {\bf Q}  {\bf A}' \right) + \frac{V}{T} - E_k'\right]  \Psi_k' &&  \ = \ 0  \ \ \ \ \ \label{eq:Schroedtime} 
\end{eqnarray}
rather then using a time-dependent density $\rho(t_j'({\bf q},t))$ as stated from \cite{vattaycomment2} in (\ref{eq:vattayerror}). The constant eigen-energies $E_k, E_k'$ are offsets in the respective Hamiltonians \cite{Duru}. The Duru--Kleinert equivalence is a fixed-energy, stationary mapping,
\[
\Psi_k({\bf q})=\Psi'_k\bigl({\bf q}'({\bf q})\bigr)
\qquad \ \ \ \ \
\frac{\partial\Psi'_k}{\partial t'}=0
\]
Consequently, no term of the form (\ref{eq:vattayerror})
\[
\frac{\partial\Psi'_k}{\partial t'}\,\nabla_{\bf q} t'({\bf q},t)
\]
occurs, contrary to what is stated in \cite{vattaycomment2} after "A rigorous mathematical examination of this transformation [sic]”. The physical time dependence is restored only afterwards,
\begin{equation}
    \psi_k({\bf q},t)=e^{-iE_k t/\hbar}\Psi_k({\bf q}) \label{eq:timewave}
\end{equation}
where \(t\) is the original physical time. Plugging (\ref{eq:timewave}) in the original Schrödinger equation (\ref{eq:SchroedCons}) shows that it is an exact solution, since
\begin{itemize}
    \item there is no $t'$ term as stated in \cite{vattaycomment2}
    \item $e^{-iE_k t/\hbar}$ implies the energy offset $E_k$ in \ref{eq:SchroedCons}
    \item the time-independent eigenwave $\Psi_k({\bf q})$ is identical $\Psi_k({\bf q}) = \Psi_k'({\bf q}'({\bf q}))$ in the original and time-scaled Schrödinger equation
\end{itemize}

Hence the only difference between the original and time-scaled wave are the eigentimes. This was detailed in the Annex of~ v3 \cite{LohmillerSlotine2026BohmPotential}, used in the original paper \cite{LohmillerSlotine2026QuantumWaves} and reproduced here at the very end of this paper. Again here \cite{LohmillerSlotine2026QuantumWaves} is using a standard result of Duru and Kleinert. 

The equivalence of (\ref{eq:SchroedCons}) and (\ref{eq:Schroedtime}) can be shown with the following Lemma \ref{lem:DeltTrafo}:
\begin{lemma}
{\it \ For a given metric ${\bf M}({\bf q}) = \frac{\partial \bar{\bf q}}{\partial {\bf q}}^T \frac{\partial \bar{\bf q}}{\partial {\bf q}}$, defining a Riemann manifold  within a $M \ge N$-dimensional orthonormal Euclidean space $\bar{\bf q}({\bf q})$, a  component wise harmonic} coordinate transformation $\bar{\bf q}({\bf q}')$ and time-scaling
\begin{eqnarray}
       ({\bf M}T)^{-1}  &=& \frac{\partial {\bf q} }{\partial {\bf q}' } \frac{\partial {\bf q} }{\partial {\bf q}' }^T  \ \ \ {\rm with} \  \ \Delta' \bar{\bf q} = 0 \   \label{eq:eikonal} \\
    t'({\bf q}, t) &=& \int_o^t T({\bf q}(\theta), \theta) d \theta \nonumber
\end{eqnarray}
always exists, where the minimal dimension of ${\bf q}'$ is $M$ or $M+1$ \cite{Nash}. 

Note that, from the Cauchy-Riemann equations, the real and imaginary components of holomorphic complex \cite{bronstejn2012taschenbuch,Riemann1851} and quaternion~\cite{Fueter1936} coordinates ${\bf q}(\bar{\bf q}({\bf q}'))$ correspond automatically to a real and harmonic coordinate transformation. If the dimension of the original position ${\bf q}$ is smaller than of ${\bf q}'$, the back mapping is reduced to a sub-manifold which maintains harmonicity.

For any given  function $f({\bf q}, t)$ equation (\ref{eq:eikonal}) implies a scalar transformation of the Laplace-Beltrami tensor operator (1.3) in \cite{LohmillerSlotine2026QuantumWaves},
\begin{eqnarray}
\Delta' f \ = \ \frac{1}{T}\  \Delta_{{\bf M}} f \nonumber
\end{eqnarray}
\label{lem:DeltTrafo}
\end{lemma}
whose tensor proof \cite{Lovelock} is given in the Appendix of \cite{lohmiller2026exactv1}. 

%In the meantime \cite{lohmiller2026exactv1} summarizes in detail the time-scaling of the continuity and Schrödinger equations, with explicit proofs that the original and time-scaled Schrödinger and continuity equation are identical for a related harmonic coordinate transformation, based on \cite{Duru}, \cite{KustaanheimoStiefel1965}, \cite{Lovelock}. A key element is here that the Laplace operator can be mapped linearly with the time-scaling $T$ from the original to time-scaled continuity and Schrödinger equation for a  related harmonic coordinate transformation \cite{lohmiller2026constrainedactionquantummechanics}:

Note that the equivalence above is only possible because the condition on the action is based on the Laplacian, a scalar, rather than on the full Hessian as in Van Vleck \cite{VanVleck1928} and Feynman \cite{Feynman}. This result allows~\cite{lohmiller2026exactv1}  to further extend applications of the approach from the nonlinear hydrogen example in \cite{LohmillerSlotine2026QuantumWaves} to the two-dimensional quartic oscillator and nonlinear pendulum. This further disproves the claims in \cite{vattaycomment2} that the approach only works for quadratic action.

%Finally the point in~\cite{vattaycomment2}, "They introduce a position-dependent time transformation to argue that the spatial derivatives of the probability density amplitude vanish for the Feynman kernel", seems to overlook that this is a standard result of~\cite{Duru}, which was already used in the original article~\cite{LohmillerSlotine2026QuantumWaves}. 

\section{Van Vleck and Feynman as Prior Art } \label{Vleck}

%, making () and () of  irrelevant. The partial differentation with respect to space is done at constant t’ , not constant t.

After a first post contradicting a standard result of Van Vleck and Feynman~\cite{Vattay2026comment},  "For bound states with spatially varying densities -- such as the harmonic oscillator (example 3.9) [...] \ the quantum potential is manifestly non-zero [sic]",  the same author's second post~\cite{vattaycomment2} now highlights that result.

However, the statement in \cite{vattaycomment2},
%"It is a rigorously proven mathematical theorem that this semi-classical trace formula is exact if and only if the Hamiltonian is at most quadratic in position and momentum [sic]", 
"The mathematical framework presented by the authors remains identical to the well-established semiclassical Van Vleck propagator [sic]" is also incorrect. Not only does it ignore the contributions in \cite{LohmillerSlotine2026QuantumWaves} as outlined in the Introduction, but it also specifically misses its highly nontrivial extension of Van Vleck and Feynman's result on the harmonic oscillator, as detailed below.

\begin{enumerate}
    \item If the action is single-valued, real and quadratic, as for a free particle or harmonic oscillator, then the propagated density is harmonic, and the solution in~\cite{LohmillerSlotine2026QuantumWaves} is the same as that derived exactly along classical paths in \cite{Feynman} and \cite{VanVleck1928}.
    
    \item If the action is multi-valued and each action branch is quadratic or linear \cite{LohmillerSlotine2026QuantumWaves}, as e.g. for a potential well, tunneling, Dirac, Pauli or Maxwell spin, then the propagated density on each branch is also harmonic. This was first introduced in \cite{LohmillerSlotine2026QuantumWaves} thanks to the exact classical branch point definition from unbounded $\Delta_{{\bf M}} \phi_j$ in Theorem 3.2, the classical quantization lemma 3.4 in \cite{LohmillerSlotine2026QuantumWaves}, and the use of complex or quaternion actions. In tunneling for instance, the relevant action cannot be real as there is no classical correspondent, but using a complex action solves the problem easily~\cite{LohmillerSlotine2026QuantumWaves}. Similarly, spin can be described using quaternion-based action~\cite{LohmillerSlotine2026QuantumWaves}, without requiring extensions of Feynman's zig-zag paths~\cite{Schulman1968PathIntegralSpin,Nielsen1988PathIntegralQuantizeSpin}.
    \item Non-quadratic action branches with position dependent $\Delta_{{\bf M}} \phi_j({\bf q}, t)$, such as the conic action branches of the two-slit experiment or the nonlinear action branches of the Aharonov-Bohm experiment \cite{LohmillerSlotine2026QuantumWaves}, or more generally the real and imaginary components of a holomorphic nonlinear complex action, can still be handled exactly since the propagated density is harmonic on each branch. This is in contrast with the semi-classical results of \cite{VanVleck1928} or \cite{Feynman}, which require linear or quadratic actions to be exact.
\end{enumerate}

Similarly, the point in~\cite{vattaycomment2}, "In their technical note, the authors acknowledge
that for non-quadratic potentials, the divergence of the classical action $\Delta_M \phi_j$ depends on spatial
coordinates. It is a rigorously proven mathematical theorem that this semi-
classical trace formula is exact if and only if the Hamiltonian is at most quadratic in position
and momentum [sic]", is incorrect:

Indeed, the Van Vleck \cite{VanVleck1928} kernel
$ \ K \approx \sum_{\gamma}
\left|
\det\!\left(
-\frac{\partial^2 \phi_\gamma}
{\partial \mathbf q_f\,\partial \mathbf q_i}
\right)
\right|^{1/2}
e^{ 
\frac{i}{\hbar}\phi_\gamma
-\frac{i\pi}{2}\mu_\gamma } $,
with $\mu_\gamma$ the Maslov index, if the propagated action $\phi_\gamma( {\bf q},t, {\bf q}_o)$ is a quadratic single-valued action, a $N \times (N -1) $ condition which implies that the pre-factor turns in a square root density. This is a much more restrictive condition than in Lemma 3.1 below, which only requires that the propagated classical square root density is harmonic. This relaxation from Van Vleck's matrix condition to a scalar condition is precisely what allows a general time scaling in Lemma 3.1 below. Finally, note that the Maslov index does not correspond to the branch points of Theorem 2.4 in \cite{LohmillerSlotine2026QuantumWaves}. In the double slit experiment, for instance, the action behind the slits is a conic function, and thus exact Feynman or Van Vleck computations based on quadratic actions cannot apply, but as we show in \cite{LohmillerSlotine2026QuantumWaves,LohmillerSlotine2026BohmPotential} $\ \Delta_{\bf M} \sqrt{\rho_j}\ $ is exactly zero.

\section{Concluding remarks}

It is appreciated that in contrast to his earlier comments \cite{Vattay2026comment} on rspa.2025.0413 \cite{LohmillerSlotine2026QuantumWaves}, answered in arXiv:2605.20443 \cite{LohmillerSlotine2026BohmPotential}, the same author of the new post \cite{vattaycomment2} does not seem to dispute any longer
that the use in \cite{LohmillerSlotine2026QuantumWaves} of tools in Feynman's \cite{Feynman}, namely
\begin{itemize}
    \item the kernel approach, which computes a wave starting at a given initial position or momentum, and is a key element in \cite{LohmillerSlotine2026QuantumWaves} to ensure that the Bohm quantum potential vanishes. This is directly verified in all examples of \cite{LohmillerSlotine2026QuantumWaves} as well as in the Pauli, Dirac and Maxwell derivations from classical action. This approach is very different from the standard Bohm-Madelung potential on the overall wave, as illustrated in detail in \cite{LohmillerSlotine2026BohmPotential} for the two-slit experiment.
    \item  for the harmonic oscillator, the kernel computation from a Taylor series expansion, which does not involve any circularity. 
\end{itemize}

However, as detailed in Sections 1, 2, 3 and 4 and summarized in the abstract, the main points of the post \cite{vattaycomment2} are again based on mis-quotations of \cite{LohmillerSlotine2026QuantumWaves} and   \cite{LohmillerSlotine2026BohmPotential}:
\begin{itemize}
  \item   In its multivalued chain-rule objection, \cite{vattaycomment2} is confusing the time $t'({\bf q}, t)$-dependent kernel computation from action in Lohmiller and Slotine \cite{LohmillerSlotine2026QuantumWaves} with the Duru and Kleinert's time-independent eigenwave equivalence of the original  Schrödinger equation with the time-scaled Schrödinger equation (16) in ~\cite{Duru}. %Indeed the rescaled time $t'$  is only used in \cite{LohmillerSlotine2026BohmPotential} \cite{LohmillerSlotine2026QuantumWaves}, for mapping the action of the Hamilton-Jacobi in $t'$ into the kernel of the Schr\"odinger equation in $t'$. 
  These results are totally decoupled.
  
  Since the $t'$-independent eigenwave is independent of time $t'$, one cannot possibly violate a time $t'({\bf q}, t)$ chain-rule. The time-dependence is removed explicitly in Duru and Kleinert's ~\cite{Duru} by adding offsets $E_k, E_k'$ to the associated Hamiltonians, which is always permissible. When the final wave is computed the offset is then included in the the standard equation $\psi_k({\bf q},t) = e^{-i E_k t / \hbar}\,\Psi_k({\bf q})$. The approach of Lohmiller and Slotine \cite{LohmillerSlotine2026QuantumWaves} is simply using this Duru and Kleinert result, making quite surprising the critic of supposedly violating a time chain-rule for a density term $\rho \bigl(t_{j}'({\bf q},t)\bigr)$ which is actually not used.
  %Our use (Lemma 3.1 and Appendix of \cite{LohmillerSlotine2026BohmPotential} in the second part of this paper, hydrogen example in \cite{LohmillerSlotine2026QuantumWaves}) of the result  (16)  of Duru and Kleinert \cite{Duru}, is incorrectly quoted that the mapping of the original to time-scaled Schrödinger equation "... is expressed as a function of this local clock $R(t_j'(x,t))$ ... [sic]". A correct quotation of (16) \cite{Duru} would have been that the time-independent quantum eigenwave of the time-scaled Schrödinger equation is mapped to the Schrödinger equation using a harmonic coordinate transformation. 

  In the meantime, \cite{lohmiller2026exactv1} now details the time-scaling of the continuity and Schrödinger equations, with explicit proofs that the original and time-scaled Schrödinger and continuity equations are identical for an associated harmonic coordinate transformation, based on \cite{Duru}, \cite{KustaanheimoStiefel1965}, \cite{Lovelock}.  The applications of the approach in \cite{LohmillerSlotine2026BohmPotential,LohmillerSlotine2026QuantumWaves} are extended in \cite{lohmiller2026exactv1} from the nonlinear hydrogen example in \cite{LohmillerSlotine2026QuantumWaves} to the two-dimensional quartic oscillator and the two-dimensional nonlinear pendulum.
\item \cite{vattaycomment2}  states incorrectly "It is a rigorously proven mathematical theorem that this semi-classical trace formula is exact if and only if the Hamiltonian is at most quadratic in position and momentum [sic]", "The mathematical framework presented by the authors remains identical to the well-established semiclassical Van Vleck propagator [sic]", by mixing the Hessian of the action with the Laplacian of the propagated density or action, by ignoring the extension of a real position to a complex or quaternion spin, and finally by ignoring the multi-valued action branches and classical quantization of \cite{LohmillerSlotine2026QuantumWaves}.

    \item  \cite{vattaycomment2} also ignores that the vast majority of the article~\cite{LohmillerSlotine2026QuantumWaves}  is decoupled from this discussion, including
    \begin{itemize}
    \item Theorem 2.4, on branch points for the classical action
    \item Lemma 3.3, which finds at the classical branch points classical randomness which explains the randomness of quantum waves
    \item Lemma 3.4, which performs a classical quantization
    \item All examples, such as the as double slit, Aharonov-Bohm, particle in the box, tunneling, harmonic oscillator, hydrogen, EPR, and QED
    \item Section 4,  which extends the approach to Klein-Gordon, Pauli, Dirac and Maxwell equations.
\end{itemize}
\end{itemize}

\clearpage

\setcounter{section}{0}

\noindent\makebox[\linewidth]{\rule{\maxdimen}{0.4pt}}

\begin{center}
{\Large \bf Interpreting Bohm-like quantum potentials in\\ "Computing quantum waves exactly from classical action" \par} \vspace{1.5em}
{\large Winfried Lohmiller and Jean-Jacques Slotine \par}
{Nonlinear Systems Laboratory \\
Massachusetts Institute of Technology \\
Cambridge, Massachusetts, 02139, USA\\
{\sl \{wslohmil, jjs\}@mit.edu} \par}
\end{center}

\begin{abstract}

The recent post arXiv:2605.02621 [14], commenting on the article "On computing quantum waves exactly from classical action" rspa.2025.0413 \cite{LohmillerSlotine2026QuantumWaves}, argues that the proof of Lemma 3.1 in \cite{LohmillerSlotine2026QuantumWaves} is missing the spatial derivative of the density, which would lead to a Bohm-like quantum potential. This technical note shows why the propagated density is harmonic in the Feynman propagator construction of Lemma 3.1. This is done by extending the proof of Lemma 3.1  explicitly with Bohm-like quantum potential terms along the stationary action paths, and then showing that these terms are exactly zero. In \cite{LohmillerSlotine2026QuantumWaves}, this property can also be verified directly on most examples (double slit, Aharonov-Bohm, potential well, harmonic oscillator, tunneling, EPR, QED), as well as in the derivations of the Pauli, Dirac, and Maxwell equations. For more general nonlinear actions, a time rescaling may be required to guarantee this space independence along stationary paths. In the hydrogen atom example, this time rescaling can be computed in closed form.

The continuity p.d.e. and the Hamilton-Jacobi p.d.e., extended by the Bohm potential, are undisputed. In contrast to the general wave of the Madelung solution \cite{Madelung1927}, Lemma 3.1 of  \cite{LohmillerSlotine2026QuantumWaves} is defined first for a propagator (more formally, a distributional kernel), and a general wave is then constructed in a second step. Recall that a Feynman propagator is a specific quantum wave, which is initialized at $t=0$ with a Dirac impulse at a given initial position or momentum. In turn, a general wave is constructed in a second step by superposing a distribution of initial conditions using the propagator.
We will see that this key difference is why the Bohm-like quantum potential terms disappear in the construction \cite{LohmillerSlotine2026QuantumWaves} (specifically, in the first step) while the  Bohm potential in the Madelung analysis does not.

This fundamental difference is also consistent with the fact that the wave construction in \cite{LohmillerSlotine2026QuantumWaves} extends naturally to relativistic contexts, while Bohmian non-locality notoriously prevents such extensions.

%\wl{Also note that  arXiv.org/abs/2605.02621 unwittingly claims to disprove a standard Feynman result on the harmonic oscillator.

\ \

\noindent Keywords $-$  Response to {\it arXiv:2605.02621}, in relation to {\it rspa.2025.0413}
"On computing quantum waves exactly from classical action", Bohm potential, Madelung equation.
\end{abstract}

 %Lohmiller-Slotine quantum wave construction 
 
\section{Introduction}

The following extends the proof of Lemma 3.1 in \cite{LohmillerSlotine2026QuantumWaves} to address the concern that a spatial derivative term is missing in the propagator calculation of Lemma 3.1. To make this issue more transparent, a derivation in fixed coordinates is used, where Bohm-like quantum potentials \cite{Bohm1952I, Bohm1952II} appear explicitly.

Lemma 3.1 in \cite{LohmillerSlotine2026QuantumWaves} computes a Feynman propagator, i.e., a wave function satisfying the Schr\"odinger equation from a given initial position ${\bf x}_o$ or momentum  ${\bf p}_o \ $, and (2.6) computes the propagated density from this initial condition. All multi-paths from this initial condition are initialized with a common density. As we shall see, this yields in turn a purely time-varying propagated density (possibly using a rescaled time for general nonlinear actions), which implies a vanishing Bohm quantum potential. Because of the specific initial Dirac density of a propagator \cite{Feynman}, this result is very different from the standard Madelung density solution \cite{Madelung1927}, which has a different initial density, action, and wave. As in \cite{Feynman}, a general wave is constructed in \cite{LohmillerSlotine2026QuantumWaves} through a second step, by {\it superposing} a distribution of initial conditions for the propagator. 

In ~\cite{LohmillerSlotine2026QuantumWaves}, the propagated density is computed explicitly across all examples and all relativistic extensions, and indeed always depends on time only. Thus the examples (double slit, Aharonov-Bohm, potential well, tunneling, harmonic potential, hydrogen atom, EPR, QED), as well as the derivations of the Pauli, Dirac, and Maxwell equations, all recover the known exact quantum results without an explicit Bohm quantum potential.

Finally, note that as in Feynman's closed-form treatment of the harmonic oscillator~\cite{Feynman}, the quantum eigenwaves (3.13) in the harmonic oscillator Example 3.9 of~\cite{LohmillerSlotine2026QuantumWaves} are obtained systematically from a Taylor expansion (equation (8.10) in~\cite{Feynman}), rather than imported a priori from existing quantum results. Also note that the statement in~\cite{Vattay2026comment} on the Bohm potential in this example contradicts the standard result in~\cite{Feynman}.

%This first release is intended as a discussion basis to align with the different stakeholders who raised a concern that the Bohm quantum potential was missed.

\section{Extended proof of Lemma 3.1}

We give an extended statement and proof of Lemma 3.1 using a Bohm-like quantum potential. The references of equations and lemmas are the same as in \cite{LohmillerSlotine2026QuantumWaves}. The discussion focuses on Lemma 3.1, and starts directly after the first paragraph of section 3 in \cite{LohmillerSlotine2026QuantumWaves}.

As we now show, the Schr\"odinger equation (1.6) can be solved on each stationary branch $j \in \mathbb{B}$ of Definition (2.2) by 
the propagator
\begin{equation}
\tag{3.1}
\psi_j({\bf x}, t, {\bf x}_o \veebar {\bf p}_o) = \sqrt{\rho_j} \ e^{\frac{i }{\hbar} \phi_j} = \sqrt{\rho_{oj}}({\bf x}_o \veebar {\bf p}_o, 0)  \ e^{- \frac{1}{2} \int_{0}^{t} \Delta_{\bf M} \phi_j({\bf x}, \theta) \ d \theta + \frac{i }{\hbar} \phi_j} \nonumber
\end{equation}
using the action field $\phi_j({\bf x}, t, {\bf x}_o \veebar {\bf p}_o)$ (2.2) and the classical propagated density field $\rho_j = \rho_{oj} \ e^{- \int_o^{t} \Delta_{\bf M} \phi_j({\bf x}, \theta) \ d \theta}$ of (2.6) in Theorem 2.4. under the constraint that the multi-path starts at ${\bf x}_o \veebar {\bf p}_o$. Replacing $\psi_j$ in the Schr\"odinger equation (1.6) yields the Madelung p.d.e. \cite{madelung_equations_wikipedia, Madelung1927} with the the continuity equation (1.3) 
\begin{equation}
  0 = \left[ \frac{\partial \phi_j}{\partial t}  + {\frac {1}{2}} \ \left( \nabla \phi_j- {\bf Q} \  {\bf A} \right)^T \ {\bf M}^{-1} \left( \nabla \phi_j- {\bf Q} \  {\bf A} \right) + V  \right] \ \psi_j + Q_j \psi_j \nonumber  %\label{eq:Madelung}
\end{equation}
%\begin{equation}
%+  \left[ \frac{\hbar}{i}  \ \frac{\partial \sqrt{\rho_j}}{\partial t}  +  \left( \frac{\hbar}{i} \nabla \phi_j  -  {\bf Q} \  {\bf A}\right) {\bf M}^{-1} \nabla \sqrt{\rho_j}  +   \frac{\hbar}{2i} \Delta_{\bf M} \Phi_j  \sqrt{\rho_j} \right] \ \psi_j / \sqrt{\rho_j} \tag{3.12}  
%\end{equation} 
The Hamilton-Jacobi equation (2.2) is augmented in the above by the Bohm-like quantum potential term 
$ \ Q_j \psi_j = - \frac{\hbar^2}{2} \ \Delta_{\bf M} \sqrt{\rho_j} \ e^{- \frac{i }{\hbar} \phi_j} \ $ \cite{Bohm1952I, Bohm1952II}.

%Let us now show that $ \ Q_j \psi_j = - \frac{\hbar^2}{2} \ \Delta_{\bf M} \sqrt{\rho_j} \ e^{- \frac{i }{\hbar} \phi_j} \ $ vanishes for the propagator (3.1). 

If $\Delta_{\bf M} \phi_j$ only depends on time and not on space, then the propagated density (2.6) assures that $ \ \Delta_{\bf M} \sqrt{\rho_j}  = 0$ and thus that $Q_j \psi_j=0$ on each stationary branch.  This is the case for the Pauli, Dirac, and Maxwell equations, as well as for all examples in the paper (double slit, Aharonov-Bohm, potential well, harmonic oscillator, tunneling, EPR, QED)  other than the hydrogen atom. 

For general nonlinear actions, $\Delta_{\bf M} \phi_j({\bf x}, t)$ {\it could indeed be space dependent}, and since $\Delta_{\bf M} \phi_j({\bf x}, t)$ is a coordinate invariant Laplace-Beltrami tensor (1.3), a pure position coordinate transformation ${\bf x}({\bf q}, t)$  will not change $\Delta_{\bf M} \phi_j({\bf x}, t) \ $ \cite{Lovelock}. D'Alembert \cite{Duru, Kleinert2009} suggested, in the course of the Kepler action computation of Example 3.10, to resolve this problem by transforming instead the {\it time} $t$ in the Hamilton-Jacobi p.d.e. (2.2) into a new time variable
$t'({\bf x},t)$ so as to make the  the associated propagated density harmonic. {\it While such a transformation can be shown to exist, in general computing it in closed form may not be trivial}. It is illustrated by the hydrogen example in \cite{LohmillerSlotine2026QuantumWaves}, and is further discussed here in the Annex based on~\cite{Duru, Kleinert2009}.

Let us summarize the result.

\begin{lemma}{\bf 3.1} \ \ \ For each branch $j$, plugging the piecewise ${\bf C}^2$ propagator $\psi_j({\bf x}, t, {\bf x}_o \veebar {\bf p}_o)$ from  (3.1) into the Schr\"odinger equation (1.6) exactly leads to the Hamilton-Jacobi p.d.e. (2.2),
 \begin{equation} 
 %\label{action_to_wave}
 \left[ \frac{\hbar}{i}  \ \frac{\partial }{\partial t} + \frac{1}{2}\left( \frac{\hbar}{i} \nabla_{\bf M} -  {\bf Q} \ {\bf A}   \right) \cdot {\bf M}^{-1} \left( \frac{\hbar}{i} \nabla - {\bf Q} \ {\bf A} \right) + V  \right] \ \psi_j =
\nonumber 
\end{equation} 
\begin{equation} 
\left[ \frac{\partial \phi_j}{\partial t}  + {\frac {1}{2}} \ \left( \nabla \phi_j- {\bf Q} \ {\bf A} \right)^T \ {\bf M}^{-1} \left( \nabla \phi_j- {\bf Q} \ {\bf A} \right) + V \right] \ \psi_j  \ = \ 0 \nonumber \tag{3.2}
\end{equation} 
The first equation is an operator equation, which becomes a product in the second equation thanks to the exponential form of (3.1). Equation (3.2) holds for piecewise ${\bf C}^2$ real, complex or quaternion actions and waves.
%Taking the sum of the waves $\psi_j$ over all branches $j$ of Definition \ref{def:branchset} then yields the overall wave $\psi$. 
\label{lem:equivalence}

This result is based on a harmonic square root density $ \ \Delta_{\bf M} \sqrt{\rho_j}  = 0 \ $ (which can be assured by  a space independent $\Delta_{\bf M} \phi_j(t)$),  derived from the action initialization (2.2) at $({\bf x}_o \veebar {\bf p}_o)$, the branch definition (2.5) and the propagated density (2.6) from $({\bf x}_o \veebar {\bf p}_o)$ of Theorem 2.4. For general nonlinear actions, a further time rescaling $ \ t'(t, {\bf x}) \ $ as in [4] (see Annex) may be needed to enforce the space independence of the associated $ \ \Delta_{\bf M} \phi_j'(t’) \ $.
\end{lemma}
The condition $ \ \Delta_{\bf M} \sqrt{\rho_j}  = 0 \ $ is fulfilled directly in all examples of this article (double slit, Aharonov-Bohm, potential well, harmonic oscillator, tunneling, EPR, QED) and in the derivations of the Pauli, Dirac, and Maxwell equations, except for the hydrogen atom example. In that example, a time-scaling is used based on equations (15) and (16) of [4] in the Annex.

In earlier attempts~\cite{Madelung1927,Schleich2013SchrdingerER, VanVleck1928, Maslov1972} to map the Hamilton-Jacobi equation to the Schr\"odinger equation, the exact equivalence (3.2) could not be shown since the specific propagated density (2.6) for a given initial ${\bf x}_o \veebar {\bf p}_o \ $, possibly in combination with a time rescaling, was not used. Instead, the Madelung density and action \cite{madelung_equations_wikipedia, Madelung1927} were used
\begin{equation}
    \rho = |\psi|^2 \ \ \ \ \ \ \ \ \ \ \ \ \phi = \hbar\,\arg(\psi) \tag{3.2*}
\end{equation}
which also solve the continuity p.d.e (1.4) and the Madelung p.d.e.~\cite{madelung_equations_wikipedia, Madelung1927} from a known wave function in polar form, $ \ \psi({\bf x},t) = \sqrt{\rho} \ e^{\frac{i}{\hbar} \phi}$. However, the propagator $\psi_j({\bf x}, t, {\bf x}_o \veebar {\bf p}_o)$ (3.1), which connects a single initial point ${\bf x}_o \veebar {\bf p}_o$ with $j \in \mathbb{B}$ multipaths to any end point ${\bf x} \in \mathbb{G}^N$ has four key differences to (3.2*),
%, compared to using $\rho({\bf x}, t) = \psi \psi^\ast$ \cite{madelung_equations_wikipedia, Madelung1927} of a single-valued action $\phi({\bf x}, t)$:
\begin{itemize}
    \item At $\ t = 0$, the path starts at ${\bf x}_o \veebar {\bf p}_o$. Assuming for instance a given initial ${\bf x}_o\ $, this implies that $\phi({\bf x}_o, 0) = 0$ and is $\phi$ undefined for ${\bf x} \ne {\bf x}_o \ $. 
    %The action is not computed from the wave. It is only computed from Hamilton-Jacobi.    
    \item At $ \ t=0$, all $j \in \mathbb{B}$ multipaths are initialized with a common density $\sqrt{\rho_{oj}}({\bf x}_o \veebar {\bf p}_o, 0)$. The initial density for $({\bf x} \ne {\bf x}_o) \veebar ({\bf p} \ne {\bf p}_o)$ is zero.
    \item The density $\rho_j$ is computed for each individual action branch $j \in \mathbb{B}$ of Theorem 2.4. 
    \item For more general nonlinear actions, a time rescaling $t'({\bf x},t)$ as in [4] (see Annex) is applied to assure that each propagated square root density is harmonic.
\end{itemize}
These differences explain why the Bohm-like  quantum potential terms \ $ \ Q_j \psi_j \ $ are  exactly zero
in Lemma 3.1, but not in general in the specific Madelung density solution (3.2*) except for a free particle.

The above allows without loss of generality to compute an overall wave $\psi({\bf x}, t)$ without a fixed initial ${\bf x}_o \veebar {\bf p}_o$ by extending the original sum over $j \in \mathbb{B}$ to the set $\mathbb{J}$, which is the same as integrating the propagator $\psi_j({\bf x}, t, {\bf x}_o \veebar {\bf p}_o)$ over all initial conditions ${\bf x}_o \veebar {\bf p}_o$. 

The separation of these two steps introduced by Feynman~\cite{Feynman}, %\wl{and \cite{Duru}}, 
starting with propagator computation and then using a sum over initial conditions, is a key element in enabling the above argumentation. It can be viewed more formally as writing the spectral decomposition of identity for the position (or momentum) operator~\cite{Schwartz1991analyse, Gelfand1964}.  It also allows  the wave construction in \cite{LohmillerSlotine2026QuantumWaves} to extend naturally to relativistic contexts, while Bohmian non-locality notoriously prevents such extensions.
%In this paper we may use interchangeably the notations $\sum$ and $\int$ in normalized sums as some indices may include mixtures of discrete and continuous quantities. [...]

All examples of the paper, as well as the derivations of the Pauli, Dirac, and Maxwell equations, are unchanged by this 
extended formulation. For more general nonlinear actions, the same caveat on the closed-form computation of $t'(t, {\bf x})$ also applies to Theorem 2, which builds on Lemma 3.1. 

\section{Example: Bohm potentials in the double slit experiment}

For illustration, let us consider for instance the Bohm-like potential terms in the first example in \cite{LohmillerSlotine2026QuantumWaves}, the double slit experiment, which Feynman deemed central to understanding quantum mechanics. Given an initial momentum $p_0$ and letting $E = \frac{p_o^2}{2 M}\ $, the action ${\phi}_j$ is 
\begin{equation}
{\phi}_j  =
\left\{ 
\begin{array}{ll}
p_o \ x^1 - E t & \text{for} \ \  x^1 < 0 \\
p_o \ r_j - E t & \text{for} \ \  x^1  \ge 0  
\end{array}
\right. \nonumber
\end{equation}
for $j =1,2 \ $, leading to the classical densities $\rho_j$ along stationary action paths
\begin{equation} 
\left\{ 
\begin{array}{ll}
\Delta_{M} \phi_j =0   & \text{  for  } x^1 < 0 \\
\Delta_{\bf M} \phi_j =\frac{2}{r_j} \frac{p_o}{M} =  \frac{2\ \dot{r}_j}{r_j}& \text{ for  } x^1 \ge 0  
\end{array}
\right.   \implies 
\rho_j = 
\left\{ 
\begin{array}{ll}
1 & \text{  for  } x^1 < 0 \\
\frac{1}{r_j^2} %= \rho_o \frac{r_o}{r_o + \frac{p_o}{M}t}  
& \text{ for  } x^1 \ge 0  
\end{array}
\right.  \implies 
\left\{ 
\begin{array}{ll}
\Delta_{M} \sqrt{\rho_j} =0 & \text{  for  } x^1 < 0 \\
\Delta_{\bf M} \sqrt{\rho_j} =0 %= \rho_o \frac{r_o}{r_o + \frac{p_o}{M}t}  
& \text{ for  } x^1 \ge 0  
\end{array}
\right. \nonumber
\end{equation}
where $\Delta_{\bf M}$ is the spherical Laplacian and the last implication uses $ \ \Delta_{\bf M} \sqrt{\rho_j} =  \frac{1}{M r} \frac{\partial^2}{\partial r^2} (r \sqrt{\rho_j}) = 0 \ $, showing that the Bohm-like terms are all exactly zero. The overall wave is thus computed from Theorem 2 as
\begin{eqnarray}
{\psi} &=&  \sum_{j \in \mathbb{B}}  \sqrt{\rho_j} \ \ e^{\frac{i }{\hbar} \phi_j}  =  e^{-\frac{i}{\hbar} E t}
\left\{ 
\begin{array}{ll}
 e^{\frac{i}{\hbar} p_o x^1 } & \ \text{for \ } x^1 < 0 \\
  \frac{1}{r_1}\  e^{\frac{i}{\hbar} p_o r_1 } + \frac{1}{r_2}\ e^{\frac{i}{\hbar} p_o r_2 }  & \ \text{for  \ } x^1 \ge 0  
\end{array}
\right. \ \ \ \ \ \  \nonumber
\end{eqnarray}
It is the same as the known quantum solution, given an initial momentum $p_0 \ $. 

The Bohm-like terms for both action branches are exactly zero, and this is what Lemma 3.1 and Theorem 2 use. By contrast, the standard Bohm potential for the overall wave is based on $|\psi|$.  A different mathematical quantity, it is in general complicated and very much non-zero. 

There is no "second step" here as the problem is stated for a given initial momentum. As noticed in \cite{LohmillerSlotine2026QuantumWaves}, Feynman's zig-zag path integrals are replaced by just two classical paths. Also note that the standard Bohm potential analysis starts with an existing wave solution, while \cite{LohmillerSlotine2026QuantumWaves} actually constructs the solution. 

Finally, note that in the numerical illustration of this Example in \cite{LohmillerSlotine2026QuantumWaves}, the ratio $ \phi / \hbar \ $ is of the order of 5, far away from the "quasi"-classical range.

\section{Summary}

Section 2 explains why the classical propagated density (4.0) does not depend on space, in contrast to a general density field. This clarifies why each Bohm-like quantum potential term \ $ \ Q_j \psi_j = - \frac{\hbar^2}{2} \Delta_{\bf M} \sqrt{\rho_j} \ e^{- \frac{i }{\hbar} \phi_j} \ $ is exactly zero in the computation of the propagator $\psi_j({\bf x}, t, {\bf x}_o \veebar {\bf p}_o)$ of Lemma 3.1 in \cite{LohmillerSlotine2026QuantumWaves}, and provides a more complete derivation of the Lemma. Section 3 illustrates the discussion on the double slit example, while the Annex discusses time rescaling in more detail.

All examples in \cite{LohmillerSlotine2026QuantumWaves}, as well as the derivations in the paper of the Pauli, Dirac, and Maxwell equations, analytically compute the classical propagated density ${\rho_j}$. Since in all cases the propagated density is either constant or only time-dependent (in the case of the hydrogen atom, after an explicit time scaling), it can be directly concluded that the Bohm-like quantum potential term $ \ Q_j  \psi_j \ $ vanishes. 
%which further confirms the correctness of all related calculations. 

Interactive simulations and visualizations of the examples can be found in the blog~\cite{Christian}.

We hope that this note clarifies the role of the Bohm-like terms in the proof of Lemma 3.1 of~\cite{LohmillerSlotine2026QuantumWaves}, and its key differences with the usual role of the Bohm quantum potential based on the Madelung density (3.2*). 

\section*{Annex: Time rescaling} \label{rescaling}

For general nonlinear potentials, a time transformation may be needed to enforce the space independence required for Lemma 3.1, as was done in the Example 3.10 of the hydrogen atom.

In this Annex, we recall how the Hamilton-Jacobi and the Schr\"odinger equation in (15) and (16) in \cite{Duru} change under a general time transformation $t'(t, {\bf x}) =\int_o^t T(\theta, {\bf x}(\theta)) d \theta $, defined by the total derivative of this time with respect to the original $t$,  
\begin{equation}
\frac{\partial t'}{\partial t} + \nabla t' \ {\bf M}^{-1}  \nabla \phi = \frac{dt'}{d t} = T \tag{4.0}
\end{equation}
The action derives from
\begin{eqnarray}
   \frac{d \phi}{dt } \ = \ L \ = \ \frac{d \phi}{dt'}\ T  \nonumber 
\end{eqnarray}
using (1.1) and (4.0). The time-scaled Hamilton Jacobi (2.2) is (see e.g., equation (15) in \cite{Duru})
\begin{equation}
       - {\frac {\partial \phi_j'}{\partial t'}} = %- {\frac {d \phi_j}{d t'}} + \nabla \phi_j \frac{ d{\bf q}}{dt'} = H'   \\
    \frac {1}{2}  \left( \nabla \phi_j' - {\bf Q}  {\bf A} \right)^T ( {\bf M} T)^{-1}  \left( \nabla \phi_j' -  {\bf Q}  {\bf A}  \right)  +  \frac{V}{T}  \tag{4.1}
\end{equation}
whose solution $ \ \phi_j'(t'(t, {\bf x}), {\bf x}) =\phi_j(t, {\bf x}) \ $ also fulfills the original Hamilton-Jacobi equation (2.2), using (4.0). 
%The spatial derivative of $\phi_j'(t', {\bf x})$ is defined on a manifold with given $t'$ and metric ${\bf M} T$, whereas the spatial derivative of $\phi_j(t, {\bf x})$ is defined on a manifold with given $t$ and metric ${\bf M}$. Hence the Laplace-Beltrami tensor (1.3) transforms between both manifolds as \cite{Lovelock}
The continuity equation is
\begin{equation}
    \frac{d }{d t'} \ln \rho_j = \Delta_{{\bf M}T} \phi_j' \tag{4.2} 
\end{equation}
with $\ d t' = T({\bf x},t) dt \ $.
%\begin{equation}
%     \Delta_{{\bf M}T} \phi_j' = \frac{1}{T}\ \Delta_{\bf M} \phi_j  - \frac{1}{2 T} \frac{\partial T}{\partial {\bf x}} ({\bf M}T)^{-1} \nabla \phi_j \tag{4.2}
%\end{equation}
Enforcing that $\Delta_{{\bf M}T} \phi_j'(t')$ in (4.1) does not depend on space is equivalent, by recursion, to imposing that a harmonic iso-$\rho(t')$ manifold increments to a harmonic iso-$\rho(t' + Tdt)$ manifold.
{\it Thus a solution $t', T$ of (4.2) assigned to each harmonic iso-$\rho(t')$ manifold always exists, although in general finding it analytically may not be trivial.}

The Schr\"odinger equation (1.6) for the Hamiltonian (4.1) is
\begin{equation}
\left[ \frac{\hbar}{i}  \ \frac{\partial }{\partial t'} + \frac{1}{2}\left( \frac{\hbar}{i} \nabla_{{\bf M}T} - {\bf Q} {\bf A} \right) \cdot ({\bf M}T)^{-1} \left( \frac{\hbar}{i} \nabla -  {\bf Q}  {\bf A} \right) + \frac{V}{T} \right]  \psi' = 0 \tag{4.4}
\end{equation}
This Schr\"odinger equation can be solved on each stationary branch $j \in \mathbb{B}$ of Definition (2.2) by 
the propagator
\begin{equation}
\psi_j'({\bf x}, t', {\bf x}_o \veebar {\bf p}_o) = \sqrt{\rho_j} \ e^{\frac{i }{\hbar} \phi_j'} = \sqrt{\rho_{oj}}({\bf x}_o \veebar {\bf p}_o, 0)   e^{- \frac{1}{2} \int_{0}^{t'_j} \Delta_{{\bf M}T} \phi_j' (\theta'_j) \ d \theta' }  e^{\frac{i }{\hbar} \phi'_j} 
\nonumber \tag{4.5}
\end{equation}
using the same argumentation as in the derivation of Lemma 3.1.

Finally, according to equation (16) in \cite{Duru}, the wave of the original Schr\"odinger equation (1.6) is given by $\psi_j({\bf x}) = \psi_j'({\bf x})$ if $\psi_j'({\bf x})$ is independent of $t'$. In this case we have to restrict to time-independent Hamiltonians $H + E_j$ in Theorem 2.4, augmented with a general energy offset $E_j$ implying a  time-independent scaling $T({\bf x})$. Accordingly, the time-scaled Schr\"odinger equation (3.2) has a stationary eigenwave solution $\psi_j'({\bf x})$, independent of $t'$. Example 3.10 of~\cite{LohmillerSlotine2026QuantumWaves} details these steps for the hydrogen atom. Incidentally, note that (16) in \cite{Duru} actually allows $\psi_j$ itself to also depend on $t$, but of course not on $t'$. As stated in the Concluding Remarks of~\cite{LohmillerSlotine2026QuantumWaves}, it is current research to find complex actions satisfying these constraints for more general nonlinear potentials.

\ \ 

\noindent {\bf Acknowledgements} \ \ We are grateful to Alexander Givental for stimulating discussions.

%suggesting to perform the proof explicitly in fixed space coordinates to clarify the role of the Bohm potential.

\normalem
\bibliographystyle{abbrv}
\bibliography{References}{}

@book{bronstejn2012taschenbuch,
  author = {Bronstein, I.N. and Semendjajew, K.A. and Musiol, G. and Mühlig, H.},
  title = {Taschenbuch der Mathematik},
  address = {Frankfurt am Main},
  year = {2012},
  publisher = {Harri Deutsch}
}

@article{Duru,
author = {Duru, I.H. and Kleinert, H.},
journal = {{Fortschritte Phys.}},
title = {{Quantum Mechanics of H-atoms from path integrals}},
year = {1982}
}

@book{Feynman,
    title = {Quantum Mechanics and Path Integrals},
    author = {Feynman, R.P. and Hibbs, A.R.},
    year = {1965},
    publisher = {McGraw-Hill}
}

@article{Fueter1936,
  author  = {Fueter, Rudolf},
  title   = {Über die analytische {D}arstellung der regulären {F}unktionen einer {Q}uaternionenvariablen},
  journal = {Commentarii Mathematici Helvetici},
  year    = {1936},
  volume  = {8},
  number  = {1},
  pages   = {371--378},
  note    = {Develops the theory of regular (Fueter-regular) quaternionic functions via the Cauchy--Riemann--Fueter equation.}
}

@book{Gelfand1964,
  title     = {Generalized Functions, Volume 4: Applications of Harmonic Analysis},
  author    = {Gelfand, I. M. and Vilenkin, N. Ya.},
  translator = {Feinstein, Amiel},
  year      = {1964},
  publisher = {Academic Press},
  address   = {New York},
  isbn      = {978-0-12-279504-6},
  series    = {Generalized Functions}
}

@book{Kleinert2009,
    title = {{Path Integrals in Quantum Mechanics, Statistics, Polymer Physics and Financial Markets}},
    author = {Kleinert, H.},
    year = {2009},
    publisher = {World Scientific}
}

@article{KustaanheimoStiefel1965,
  author  = {Kustaanheimo, P. and Stiefel, E.},
  title   = {Perturbation Theory of Kepler Motion Based on Spinor Regularization},
  journal = {Journal f{\"u}r die reine und angewandte Mathematik},
  volume  = {218},
  pages   = {204--219},
  year    = {1965}
}

@article{Bohm1952I,
  author = {Bohm, David},
  title = {A Suggested Interpretation of the Quantum Theory in Terms of ``Hidden'' Variables, {I}},
  journal = {Physical Review},
  volume = {85},
  number = {2},
  pages = {166--179},
  year = {1952},
  month = {jul},
  doi = {10.1103/PhysRev.85.166},
  publisher = {American Physical Society}
}

@article{Bohm1952II,
  author = {Bohm, David},
  title = {A Suggested Interpretation of the Quantum Theory in Terms of ``Hidden'' Variables, {II}},
  journal = {Physical Review},
  volume = {85},
  number = {2},
  pages = {180--192},
  year = {1952},
  month = {jul},
  doi = {10.1103/PhysRev.85.180},
  publisher = {American Physical Society}
}

@article{LohmillerSlotine2026QuantumWaves,
  author  = {Lohmiller, Winfried and Slotine, Jean-Jacques E.},
  title   = {On computing quantum waves exactly from classical action},
  journal = {Proceedings of the Royal Society A},
  volume  = {482},
  number  = {2336},
  pages   = {20250413},
  year    = {2026},
  month = {April},
  doi     = {10.1098/rspa.2025.0413}
}

@article{lohmiller2026exactv1,
  author        = {Winfried Lohmiller and Jean-Jacques Slotine},
  title         = {Exact Computation of Quantum Wave Functions for Nonlinear Potentials},
  year          = {2026},
  month         = sep,
  eprint        = {2609.14480},
  archiveprefix = {arXiv},
  primaryclass  = {quant-ph},
  url           = {https://arxiv.org/abs/2609.14480},
  note          = {arXiv:2609.14480v1 [quant-ph], version 1}
}

@misc{LohmillerSlotine2026BohmPotential,
  author    = {Lohmiller, Winfried and Slotine, Jean-Jacques},
  title     = {{Interpreting Bohm-like Quantum Potentials in "Computing Quantum Waves Exactly from Classical Action"}},
  year      = {2026},
  eprint    = {2605.20443},
  archivePrefix = {arXiv},
    month     = {May},
  doi       = {10.48550/arXiv.2605.20443},
  url       = {https://arxiv.org/abs/2605.20443}
}

@book{Lovelock,
    title = {Tensors, Differential Forms, and Variational Principles},
    author = {Lovelock, D. and Rund, H.},
    year = {1989},
    publisher = {Dover}
}

@article{Madelung1927,
  author    = {E. Madelung},
  title     = {{Quantentheorie in hydrodynamischer Form}},
  journal   = {Zeitschrift f{\"u}r Physik},
  volume    = {40},
  number    = {3-4},
  year      = {1927},
  doi       = {10.1007/BF01400372}
}

@misc{madelung_equations_wikipedia,
  title        = {Madelung equations},
  author       = {{Wikipedia contributors}},
  howpublished = {\url{https://en.wikipedia.org/wiki/Madelung_equations}}
}

@book{Maslov1972,
  author = {Victor P. Maslov},
  title = {Théorie des perturbations et méthodes asymptotiques},
  publisher = {Éditions Mir},
  year = {1972},
}

@article{Pauli,
    author = {Pauli, W.},
    journal = {{Zeitschrift für Physik}},
    title = {{Zur Quantenmechanik des magnetischen Elektron}},
    year = {1927}
}

@article{Schulman1968PathIntegralSpin,
  title   = {A Path Integral for Spin},
  author  = {Schulman, Lawrence S.},
  journal = {Physical Review},
  volume  = {176},
  pages   = {1558--1569},
  year    = {1968},
  doi     = {10.1103/PhysRev.176.1558}
}

@article{Nielsen1988PathIntegralQuantizeSpin,
  title   = {A Path Integral to Quantize Spin},
  author  = {Nielsen, Holger Bech and Rohrlich, Daniel},
  journal = {Nuclear Physics B},
  volume  = {299},
  pages   = {471--483},
  year    = {1988},
  doi     = {10.1016/0550-3213(88)90545-7},
}

@article{Nash,
 title = {{$C^1$ Isometric Embeddings}},
  author = {Nash, John},
 year = {1954},
journal = {{Annals of Mathematics. 2nd Series, 60 (3)}}
}

@phdthesis{Riemann1851,
  author      = {Riemann, Bernhard},
  title       = {Grundlagen f\"{u}r eine allgemeine Theorie der Functionen einer ver\"{a}nderlichen complexen Gr\"{o}sse},
  school      = {Universit\"{a}t G\"{o}ttingen},
  year        = {1851},
  address     = {G\"{o}ttingen},
}

@article{Schleich2013SchrdingerER,
  title = {Schr{\"o}dinger equation revisited},
  author = {Wolfgang P. Schleich and Daniel M. Greenberger and Donald H. Kobe and Marlan O. Scully},
  journal = {P.N.A.S.},
  year = {2013},
  volume = {110},
  number = {14},
  pages = {5374--5379},
  doi = {10.1073/pnas.1302475110},
  url = {https://www.pnas.org/doi/10.1073/pnas.1302475110}
}

@book{Schwartz1991analyse,
  author    = {Schwartz, Laurent},
  title     = {Analyse},
  publisher = {Hermann},
  year      = {1991},
  address   = {Paris},
  language  = {French},
  pages     = {4 v.},
  isbn      = {2705659005},
}

@article{VanVleck1928,
  author = {John H. Van Vleck},
  title = {The Correspondence Principle in the Statistical Interpretation of Quantum Mechanics},
  journal = {P.N.A.S},
  volume = {14},
  number = {2},
  pages = {178--188},
  year = {1928},
  doi = {10.1073/pnas.14.2.178},
  url = {https://www.pnas.org/doi/10.1073/pnas.14.2.178},
}

@misc{Vattay2026comment,
  author       = {Vattay, G{\'a}bor},
  title        = {{C}omment on `{O}n computing quantum waves exactly from classical action', arXiv:2605.02621 [quant-ph]},
  year         = {2026},
  eprint       = {2605.02621},
  archiveprefix= {arXiv},
  url          = {https://arxiv.org/abs/2605.02621},
  doi          = {10.48550/arXiv.2605.02621},
}

@misc{vattaycomment2,
  title        = {Reply to "{I}nterpreting {B}ohm quantum potentials in `{C}omputing quantum waves exactly from classical action'", ar{X}iv:2606.05197},
  author       = {Vattay, Gabor},
  journal      = {arXiv preprint arXiv:2606.05197},
   year         = {2026},
  eprint       = {2606.05197},
  archivePrefix= {arXiv},
  primaryClass = {quant-ph},
  doi          = {10.48550/arXiv.2606.05197}
}

@misc{Christian,
  author       = {Pehle, Christian},
  title        = {Quantum waves from classical branches — interactive simulations},
  howpublished = {\url{https://cpehle.github.io/blog/classic_quantum_branch_sums.html}}
}

\end{document}